\documentclass[twocolumn,amsmath,amssymb,aps,prb,floatfix,superscriptaddress,longbibliography]{revtex4-1}

\usepackage{float}
\floatstyle{boxed}

\newcommand{\bra}[1]{\langle{#1}\vert}
\newcommand{\ket}[1]{\vert{#1}\rangle}
\newcommand{\avg}[1]{\langle {#1} \rangle}

\newcommand{\Tr}{\text{Tr}}

\usepackage{color}
\usepackage{graphicx}
\usepackage{dcolumn}
\usepackage{bm}
\usepackage[utf8]{inputenc}
\usepackage[T1]{fontenc}
\usepackage[dvipdfx,cmyk]{xcolor}     
\usepackage[colorlinks,linkcolor=red,citecolor=brown,urlcolor=blue]{hyperref}

\begin{document}

\title{Joint estimation of phase and phase diffusion for quantum metrology}

\author{Mihai-Dorian Vidrighin}
\affiliation{QOLS, Blackett Laboratory, Imperial College London, London SW7 2BW, UK}
\affiliation{Clarendon Laboratory, Department of Physics, University of Oxford, OX1 3PU, United Kingdom}
\author{Gaia Donati}
\affiliation{Clarendon Laboratory, Department of Physics, University of Oxford, OX1 3PU, United Kingdom}
\author{Marco G. Genoni}
\affiliation{QOLS, Blackett Laboratory, Imperial College London, London SW7 2BW, UK}
\affiliation{Department of Physics \& Astronomy, University College London, Gower Street, London WC1E 6BT, United Kingdom}
\author{Xian-Min Jin}
\affiliation{Clarendon Laboratory, Department of Physics, University of Oxford, OX1 3PU, United Kingdom}
\affiliation{Department of Physics and Astronomy, Shanghai Jiao Tong University, Shanghai 200240, PR China}
\author{W. Steven Kolthammer}
\affiliation{Clarendon Laboratory, Department of Physics, University of Oxford, OX1 3PU, United Kingdom}
\author{M.S. Kim}
\affiliation{QOLS, Blackett Laboratory, Imperial College London, London SW7 2BW, UK}
\author{Animesh Datta}
\affiliation{Clarendon Laboratory, Department of Physics, University of Oxford, OX1 3PU, United Kingdom}
\author{Marco Barbieri}
\affiliation{Clarendon Laboratory, Department of Physics, University of Oxford, OX1 3PU, United Kingdom}
\author{Ian A. Walmsley}
\affiliation{Clarendon Laboratory, Department of Physics, University of Oxford, OX1 3PU, United Kingdom}

\begin{abstract}

Phase estimation, at the heart of many quantum metrology and communication schemes, can be strongly affected by noise, whose amplitude may not be known, or might be subject to drift. Here, we investigate the joint estimation of a phase shift and the amplitude of phase diffusion, at the quantum limit. For several relevant instances, this multiparameter estimation problem can be effectively reshaped as a two-dimensional Hilbert space model, encompassing the description of an interferometer phase probed with relevant quantum states -- split single-photons, coherent states or \text{N}00\text{N} states. For these cases, we obtain a trade-off bound on the statistical variances for the joint estimation of phase and phase diffusion, as well as optimum measurement schemes. We use this bound to quantify the effectiveness of an actual experimental setup for joint parameter estimation for polarimetry. We conclude by discussing the form of the trade-off relations for more general states and measurements. 
\end{abstract}

\maketitle

Efficient sensing and imaging of samples that cannot be exposed to high field intensities require the optimisation of the amount of information acquired from each run of an experiment. Judiciously designed quantum strategies can lead to significant improvements of sensitivities, when compared to classical strategies with probes of equivalent energy content~\cite{metrology}. Phase estimation illustrates well the advantages of quantum metrology, with wide-ranging practical applications~\cite{HB,weinland,revmet, exp2, exp3, exp4}. Variations of many physical properties such as weak fields \cite{smallfield}, displacements, or changes in concentration \cite{concentration1, concentration2} can be efficiently observed as phase shifts.


A central aspect of sensing in a real scenario is the interaction between the system and the environment. When one takes into account this coupling, the promised quantum enhancement is likely to be lost \cite{noisyMetro}. This has been extensively shown in the case of a lossy interferometer
\cite{loss2,loss3,loss1,loss4} while, more recently, theoretical and experimental efforts have been directed at studying the limits of phase estimation
in the presence of phase diffusion~\cite{genoni1, genoni2, genoni3, Escher,durkin}.
Phase-diffusive noise describes fluctuations in the modes
of an interferometer, which can be modelled as a random phase-kick. This process reduces the visibility of interference, directly affecting precision measurements as, for instance, in Ramsey interferometry.
The works cited above studied the limits of phase estimation given a known amount of phase diffusion.

However, in several physical processes, such as path length fluctuations of a stabilised interferometer, thermal fluctuations of an optical fibre and weak coupling of the probed system to the environment, phase and phase diffusion may vary in time. Consequently, phase estimation relying on past estimates of the magnitude of phase diffusion may lead to inaccuracies.
A more accurate solution consists in estimating the phase shift and the phase diffusion in a simultaneous, joint scheme. This allows for the monitoring of both parameters on the relevant time-scale avoiding systematic effects. Moreover, the consideration of joint estimation for phase and noise amplitude leads to a fair accounting of resources, eliminating the need for precise calibration before the estimation. These represent important motivations for exploring the fundamental limits of this multiparameter scenario. Similar investigations have received considerable attention recently\cite{multi1, multi2, multi3, multi4, multi5, multi6, multi7, multi8, multi9, spins, guta, pete}.

In this paper, we study theoretically as well as experimentally the simultaneous estimation of phase shift and diffusion using quantum states that can be effectively described in a two-dimensional Hilbert space (as qubits). This provides a description of two-arm interferometry with some of the relevant probe states in quantum metrology -- coherent and $\text{N}00\text{N}$ states \cite{noonAndSagnac}. We find that the quantum precision limit for the estimation of each of the parameters cannot be reached for both parameters simultaneously. We derive a trade-off relation that must be obeyed by the statistical variances attainable through any physical measurement and use it to identify a double-homodyne setup as the optimal choice for phase and phase diffusion measurements for the considered set of probe states. We show this trade-off via a tunable measurement, which we characterize by quantum detector tomography. Finally, we analyse the applicability of our bounds for more general states and measurements. To this end, we present the results of numerical searches for optimum measurements acting on Holland-Burnett states \cite{HB}, as well as on a pair of qubits. We also present the performance of the double-homodyne measurement setup which we identified as optimal, in the presence of losses.

\section*{Results}

{\bf Quantum estimation theory} 
Fisher information (FI) provides an asymptotic measure of the amount of information on the parameters of a system that is acquired by performing a measurement on it. For parameters $\boldsymbol{\lambda}$, in terms of the probabilities associated to the measurement results $p_n=p_n(\boldsymbol{\lambda})$, the FI matrix elements read\cite{CRB} $F_{i,j}=\sum_{n} p_n\frac{\partial^2}{\partial \lambda_i \partial \lambda_j}\log(p_n)=\sum_{n} \frac{1}{p_n}\frac{\partial p_n}{\partial \lambda_i}\frac{\partial p_n}{\partial \lambda_j}$. The Cram\'er-Rao bound states that, for all unbiased estimators $\hat{\boldsymbol{\lambda}}$, the expected covariance matrix with elements defined as $\gamma_{i,j}=\avg{\hat{\lambda}_i \hat{\lambda}_j}-\avg{\hat{\lambda}_i}\avg{\hat{\lambda}_j}$ satisfies
\begin{equation}
\label{CRbound}
\boldsymbol{\gamma}\geq(M \text{\bf F})^{-1}
\end{equation}
where $M$ is the number of experimental runs. The Cram\'er-Rao bound is saturated asymptotically by the maximum likelihood estimator,
upon elimination of systematic errors. FI provides a useful indicator of the optimality of a given experiment and constitutes a useful tool for designing measurements with the goal of minimising statistical errors.

For a set of probabilities originated from measurements on a quantum system, the ultimate limit on the covariance matrix is set by the quantum Cram\'er-Rao bound in terms of the quantum Fisher information (QFI) matrix \cite{HelstromBook,MatteoIJQI}. Introducing the symmetric logarithmic derivative  (SLD) operator  $L_j$ for parameter $\lambda_j$, obeying $2 \partial_{\lambda_j} \rho=L_j\rho+\rho L_j$, the QFI matrix is defined as $H_{ij}=\hbox{Re}[\hbox{Tr}[\rho L_i L_j]]$ and it bounds the FI matrix corresponding to any particular measurement: $\text{\bf H}\geq \text{\bf F}$.
For a single parameter, the ultimate bound can always be achieved choosing the measurement given by the eigenvectors of the SLD operator. In the case of a multi-parameter problem, if the SLDs corresponding to different parameters do not commute then the FI values for the two parameters are maximized by incompatible measurements.

{\bf Interferometry with phase diffusion} 
We consider an interferometer with phase difference $\phi$ between its two arms. The annihilation operators corresponding to each arm are labeled $\hat{a}$ and $\hat{b}$.
Different physical processes lead to phase diffusion and the corresponding channel can be modelled as a random phase shift distributed according to a normal distribution of width $\Delta$, called the noise amplitude. Acting on a mode $a$ with initial state $\rho_{in}$, the phase diffusion channel yields
\begin{equation}
\rho = \mathcal{N}_{\Delta}(\rho_{in})=\frac{1}{\sqrt{2\pi}\Delta}\int\!\!\text{d}\xi\,\,e^{-\frac{\xi^2}{2\Delta^2}} U_{\xi}\rho_{in} U_{\xi}^{\dagger}
\end{equation}
where $U_{\xi}=\exp(i \xi \hat{a}^{\dagger}\hat{a})$ is the phase shift operator. In the Fock basis, the result is the exponential erasing of the off-diagonal elements of the density matrix:
\begin{equation}
\mathcal{N}_{\Delta}(\ket{n}\bra{m})=e^{-\Delta^2(n-m)^2}\ket{n}\bra{m}.
\end{equation}
This mapping can be attained, alternatively, by solving the master equation corresponding to phase diffusion \cite{genoni1}.
Quantum strategies aiming at an enhancement of the precision in phase estimation make use of
$\text{N}00\text{N}$ states, defined as $\frac{1}{\sqrt{2}}\left(|N0\rangle + |0N\rangle \right)$. Even under phase diffusion, the evolution of these states lies in the two-dimensional space spanned by  $\ket{N,0}{=}\frac{1}{\sqrt{N!}}(\hat a^\dagger)^N\ket{00}$ and $\ket{0,N}{=}\frac{1}{\sqrt{N!}}(\hat b^\dagger)^N\ket{00}$.
A two-dimensional picture also describes classical phase estimation strategies relying on coherent states. Indeed, a coherent state with amplitude $\alpha$ yields the same precision as a collection of $|\alpha|^2$ independent single photons ({\em i.e.} $\text{N}00\text{N}$ states with $N{=}1$).
In these relevant cases, our two-mode probe state can be effectively modelled as a single qubit 
\begin{equation}
\rho_0=\left(\begin{array}{cc}
\cos^2(\frac{\theta}{2}) & \cos(\frac{\theta}{2})\sin(\frac{\theta}{2})\\
\cos(\frac{\theta}{2})\sin(\frac{\theta}{2}) & \sin^2(\frac{\theta}{2})
\end{array}\right),
\end{equation}
which, acted upon by a phase shift $\phi$ and a phase diffusion channel parametrized by $\Delta$, yields
\begin{equation}
\rho=\left(\begin{array}{cc}
\cos^2(\frac{\theta}{2}) & \cos(\frac{\theta}{2})\sin(\frac{\theta}{2})e^{-\text{i} \phi-\Delta^2}\\
\cos(\frac{\theta}{2})\sin(\frac{\theta}{2})e^{\text{i} \phi-\Delta^2} & \sin^2(\frac{\theta}{2})
\end{array}\right).
\label{2dState}
\end{equation}
The QFI matrix corresponding to parameters $\phi$ and $\Delta$, depending on the probe parameter
$\theta$ can be calculated, using the SLD.
\begin{equation}
\text{\bf H}_\theta(\phi,\Delta)=\text{\bf H}_\theta(\Delta)=\sin^2\theta\left(\begin{array}{cc}
e^{-2 \Delta^2} & 0\\
0 & \frac{4\Delta^2}{e^{2\Delta^2}-1}
\end{array}\right).
\end{equation}
The maximum QFI corresponds to equatorial states with $\theta=\pi/2$. From now on we shall refer to the diagonal elements of the matrix $\text{\bf H}_{\pi/2}(\Delta)$ as $H_{11}$ and $H_{22}$.
For $\text{N}00\text{N}$ states and for coherent states with amplitude $\alpha$, the QFI matrices read
\begin{align}
\text{\bf H}^{(\text{N}00\text{N})}(\Delta) &=N^2 \text{\bf H}_{\pi/2}( N\Delta)\:, \\
\text{\bf H}^{(coh)}(\Delta) &=|\alpha|^2 \text{\bf H}_\theta(\Delta) ,
\end{align}
respectively.

The SLDs corresponding to the two parameters do not commute. However for equatorial states (corresponding to balanced interferometers), the expectation value of their commutator vanishes, {\em i.e.} ${\Tr[\rho(L_1 L_2-L_2 L_1)]=0}$ for $\theta=\pi/2$. In principle, when this condition is satisfied, a measurement that attains the QFI for joint estimation of both parameters can be constructed\cite{guta}.
This requires a collective measurement on multiple copies of evolved probe states, which is a challenging task to implement. Therefore, we firstly restrict our search for an optimal strategy to separable positive operator valued measurements (POVMs) i.e. measurements that act on probe states individually. We discuss extensions to joint measurements subsequently.

{\bf Trade-off in the estimation precision for $\phi$ and $\Delta$}
In order to assess the performances of these measurements we consider the quantities  $F_{1,1}/H_{1,1}$ and $F_{2,2}/H_{2,2}$, {\em i.e.} the ratios between the FI and QFI values for $\phi$ and $\Delta$.
Finding a relation between these ratios would effectively express the interplay that exists between the estimator variances corresponding to the two parameters.
As shown in the Supplementary Methods, a trade-off relation can be derived, which is obeyed for all probe states and separable measurements:
\begin{equation}
\label{rel}
\frac{F_{1,1}}{H_{1,1}}+\frac{F_{2,2}}{H_{2,2}}\leq  1\:.
\end{equation}
The most naive bound for the quantity in Equation \ref{rel} is equal to 2, and it would be in principle achievable by means of a measurement which is optimal for both parameters. With this inequality, we not only prove that such measurement does not exist, but we also quantify the maximum precision achievable in a joint estimation. Specifically, we prove that any measurement that is independently optimum for the estimation of one of the parameters is completely insensitive to the other. This bound is saturated by all POVMs with elements in the equatorial plane of the Bloch sphere which have the form
\begin{equation}
\label{chi}
\Pi_j=\frac{n_j}{2}\left(\begin{array}{cc}
\frac{1}{2} & \frac{1}{2} e^{-\text{i} \chi_j}\\
 \frac{1}{2} e^{\text{i} \chi_j} & \frac{1}{2}
\end{array}\right),
\end{equation}
with $0{<}n_j{<1}$, $0{\leq}\chi_j{\leq}2\pi$, where the probability of outcome $j$ is $\hbox{Tr}[\rho \Pi_j]$ and $\sum_j \Pi_j=\mathbb{I}$.

A further bound on statistical variances can be derived from this relation and Equation \ref{CRbound}. The expected variance of the phase shift estimator obeys $\gamma_{1,1}=\text{Var}(\phi) \geq [M (F_{1,1}-F_{1,2}^2/F_{2,2})]^{-1}$ and an analogous relation can be written for the phase diffusion amplitude. Using the fact that the off-diagonal elements of the FI matrix are real numbers, we get $\text{Var}(\phi)\geq(M F_{1,1})^{-1}$ and $\gamma_{2,2}=\text{Var}(\Delta)\geq (M F_{2,2})^{-1}$. Notice that the off-diagonal elements of the FI matrix correspond to the coupling of estimators for the two parameters, which results in increased statistical errors. Thus, the statistical variances obey
\begin{equation}
\label{rel1}
\frac{H_{1,1}^{-1}}{\text{Var}(\phi)}+\frac{H_{2,2}^{-1}}{\text{Var}(\Delta)}\leq  M.
\end{equation}

This inequality is one of our main results. It is saturated when the inequality given in Equation \ref{rel} is saturated and the off-diagonal elements of the FI matrix are zero.

{\bf An optimal measurement} 
In the Supplementary Methods, we show that the bound in Equation \ref{rel1} can be saturated for POVMs in the equatorial plane that are symmetric with respect to the measured state -- meaning that for each operator of the form given in Equation \ref{chi}, parametrized by $n_j=n$ and $\chi_j=\phi+\delta$, the POVM set contains another element, parametrized by $n_{j'}=n$ and $\chi_{j'}=\phi-\delta$ for some $\delta$. Note that in general the POVM saturating the bound depends on the specific value of the phase $\phi$.

We prove, also in the Supplementary Methods that a double homodyne setup, combining modes $a$ and $b$ on a beam splitter and measuring the $X$ and $P$ quadratures, respectively, in the beam splitter's two outputs, saturates the bound in Equation \ref{rel1} independently on the value of $\phi$. Figure \ref{homodynes} shows the dependence of the variances on $\Delta$ for this setup, which is depicted in Supplementary Fig. 1.

\begin{figure}[H]
\centering
\includegraphics[width=0.475\textwidth]{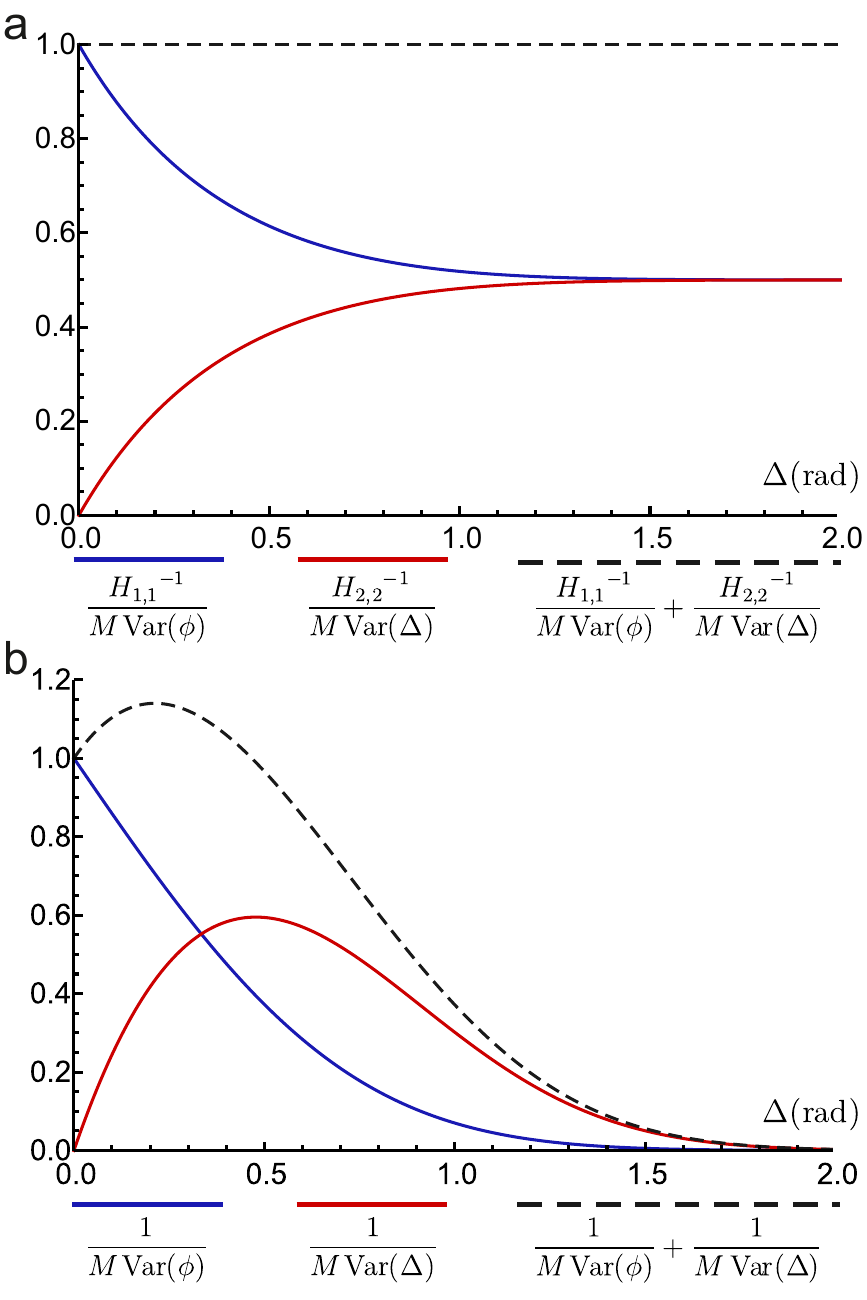}
\caption{\textbf{Joint parameter estimation by double homodyne.} (a) The ratios between optimum single parameter statistical variances and statistical variances that can be achieved with the double homodyne setup, as a function of phase diffusion amplitude, for a split single photon probe state  (first order \text{N}00\text{N} state). Note that, for low phase diffusion amplitude, the homodyne setup measures phase optimally. (b) FI elements, for a split single photon. Blue corresponds to phase estimation precision and red to phase diffusion amplitude estimation precision, while black corresponds to the sum of the two. For \text{N}00\text{N} states with N photons, plots a and b scale by a factor of $N^2$ vertically and by a factor of $1/N$ horizontally. A schematic of this measurement setup, which implements a continuous measurement, is included as Supplementary Fig. 1. }
\label{homodynes}
\end{figure}

\begin{figure}[H]
\centering
\includegraphics[width=0.5\textwidth]{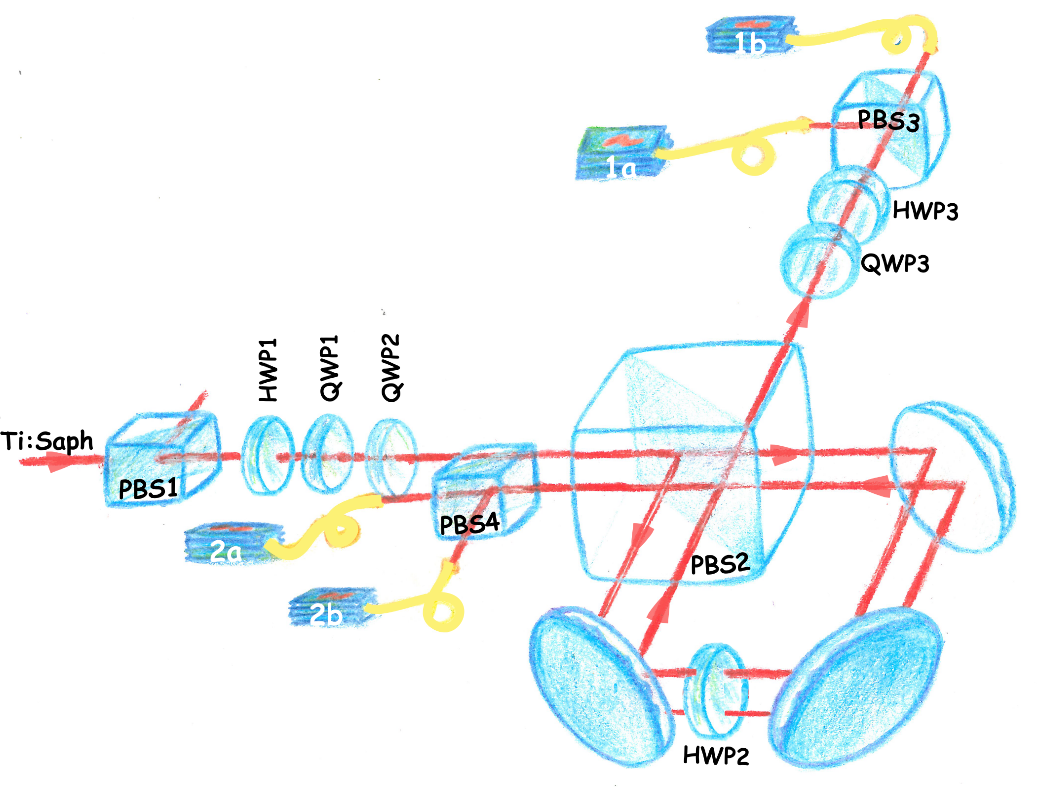}
\caption{\textbf{Experimental setup.} In our setup, the two modes $a$ and $b$ correspond to the horizontal ($H$) and vertical ($V$) polarisation of a single spatial mode. 
In this basis, the diagonal polarisation states are defined as $\ket{D}=\frac{1}{\sqrt{2}}(\ket{H}+\ket{V})$ and $\ket{A}=\frac{1}{\sqrt{2}}(\ket{H}-\ket{V})$ and the circular polarisation states as $\ket{R}=\ket{H}+\text{i}\ket{V}$ and $\ket{L}=\ket{H}-\text{i}\ket{V}$. The four outputs of the polarisation interferometer correspond approximately to the following POVM operators acting on the input polarisation state: $\{\Pi_{1a}=k\ket{D}\bra{D}, \Pi_{1b}=k\ket{A}\bra{A}, \Pi_{2a}=(1-k)\ket{R}\bra{R}, \Pi_{2b}=(1-k)\ket{L}\bra{L}\}$, where $k$ is a tunable parameter. This measurement should saturate the inequality given in Equation \ref{rel1} for certain input states. PBS -- polarizing beam splitter; HWP -- half-wave plate; QWP -- quarter-wave plate.}
\label{setup}
\end{figure}

\begin{figure}[H]
\centering
\includegraphics[width=0.475\textwidth]{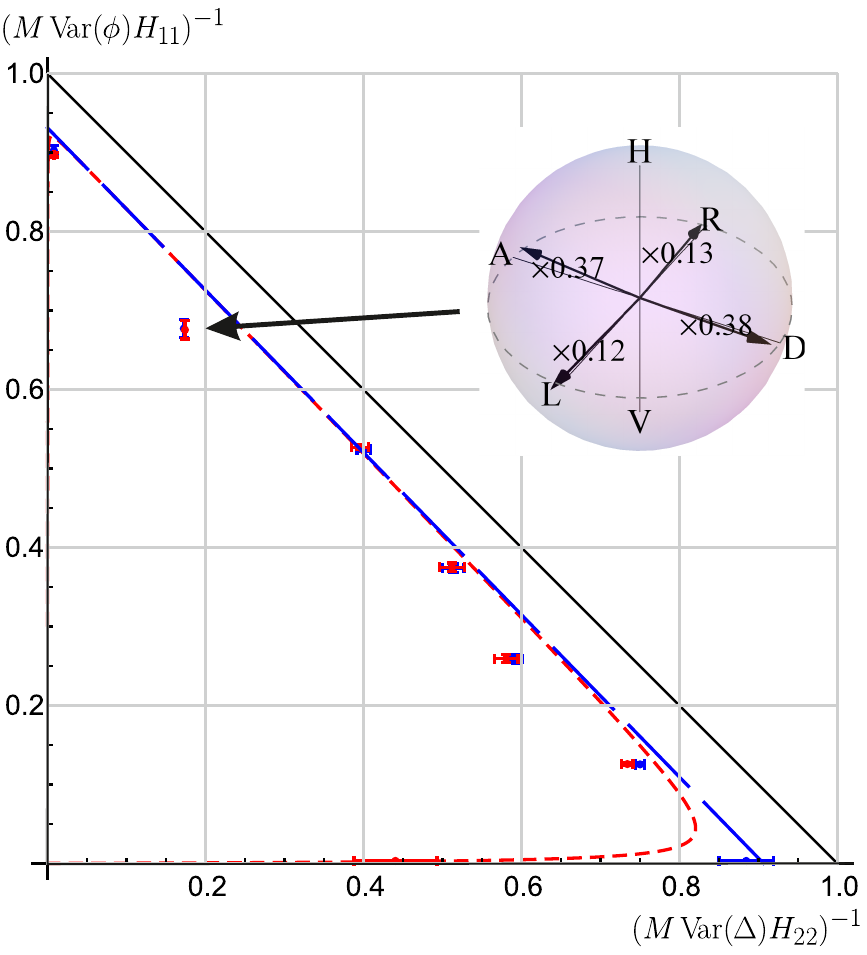}
\caption{
\textbf{Experimental results.} Parametric plot of the estimates for the ratios between optimum single parameter statistical variances and optimum statistical variance, achievable with our experimental setup (details in the Methods section). Blue points are calculated for a phase shift of approximately $\frac{\pi}{2}$, optimized to obtain null off-diagonal elements of the FI matrix. Red points are calculated for a phase shift differing by $1^{\circ}$ with respect to the one corresponding to the blue points. Error bars represent twice the standard deviation, obtained by Monte-Carlo simulation (details in the Supplementary Methods). The black line gives the ultimate limit. The blue and red dashed lines give the theoretical prediction for the blue and red points, respectively, assuming visibilities of 96.5\% for outputs 1a and 1b and 99.4\% for outputs 2a and 2b. Insert: Bloch sphere representation of the estimated POVM operators, for the indicated setting (all the estimated POVMs are represented in Supplementary Fig. 2).  The vectors represent the measurement operators corresponding to the 4 outputs, each normalized. The numbers written on the vectors are the trace norms of the corresponding operators, weighed by the total trace of the 4 operators. We choose a phase diffusion amplitude $\Delta=0.25 \,\text{rad}$ ($\approx 14^\circ$). Details on how this figure is obtained are presented in the Supplementary Methods.}
\label{exp}
\end{figure}

{\bf Experiment}
We adopt our theory to quantify the effectiveness of an actual experimental setup for joint parameter estimation for polarimetry by investigating how close the implementation compares with the optimal bound in Eq. (11). We are not aiming at demonstrating a quantum advantage and realise our implementation with coherent states. The joint estimation of phase and phase diffusion requires a measurement with at least three outputs. This is because the FI matrix corresponding to any single qubit (two-output) projective measurement is singular. Thus, it cannot be inverted, yielding unbounded estimator variances. We implemented a four-outcome measurement based on a displaced Sagnac polarisation interferometer \cite{noonAndSagnac, Mosley_2006}, depicted in Figure \ref{setup}. Our measurement realises a mixture of the optimal projective measurements for estimating the phase and the one for the phase diffusion amplitude. The setup can be arranged to tune the different weights of these measurements by rotating a waveplate.
Complete information about the POVM associated to each of these measurements is obtained via detector tomography \cite{detectortomo}. This technique adopts a quorum of input states and records the probabilities of the outcomes. The Born rule then allows to reconstruct the measurement operator. These reconstructed POVMs are then used to compute the relevant FI matrix, as detailed in the Methods section. In Figure \ref{exp} we report our results, where the variances Var$(\phi)$ and Var$(\Delta)$ have been estimated from the classical Cram\'er-Rao bound. The plot is obtained by varying the measurement from $\sigma_x$, the optimal measurement for phase estimation, to $\sigma_y$, the optimal measurement for estimating the diffusion amplitude.
The experimental results are close to the optimum precision given by Equation \ref{rel1}, with the main imperfection of the implementation stemming from non-unit interference visibility and imperfect alignment of the setup. The precision for the estimates of $\phi$ depends strongly on the measurement visibility corresponding to outputs $1a$ and $1b$ (according to Figure \ref{exp}). For $\Delta$, the precision strongly depends on the visibility corresponding to outputs $2a$ and $2b$. The influence of non-unit visibility is more pronounced for the latter, as we detail in the Supplementary Methods.


{\bf Extensions}. We have so far restricted both theoretical and experimental studies to measurements on single quantum probes. Collective measurements on multiple copies of probe states may get closer to the multiparameter quantum Cram\'er-Rao bound in some cases~\cite{guta}. We study this for the simplest nontrivial case, that of an entangled projective measurement on a pair of qubit probe states that have undergone the same phase shift and phase diffusion. In the Supplementary Methods, we analyze the performance of a Bell measurement (in the basis in which the states of Equation ~(\ref{2dState}) are written).
\begin{figure}[H]
\centering
\includegraphics[width=0.475\textwidth]{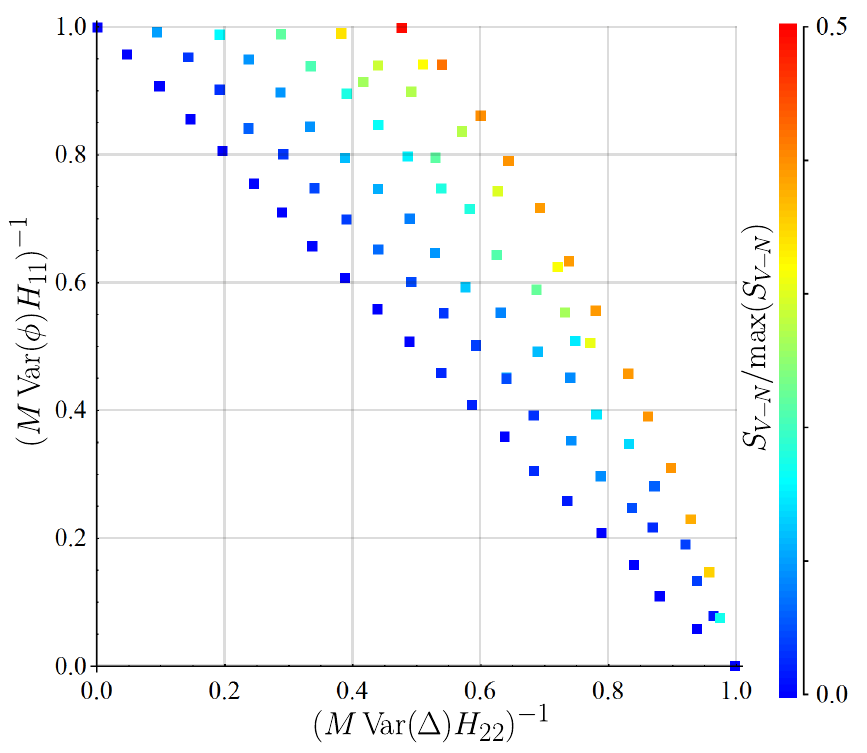}
\caption{\textbf{Collective measurements.}
Results of a simulated annealing search over all projectors acting on the space of two qubit probe states. Points are shown for all coordinates that signify a violation of the bound given by Equation ~(\ref{rel1}). Color illustrates the smallest value of total entropy of entanglement of the projectors found by the search, weighed by the maximum possible value (which corresponds to a Bell measurement). The maximum sum of coordinates in this graph is $1.48$ and the corresponding entropy of entanglement is $0.425$. The search is performed for a phase diffusion amplitude $\Delta=0.25 \,\text{rad}$ ($\approx 14^\circ$). The details of the search are presented in the Methods section and Supplementary Methods.}
\label{collectmeas}
\end{figure}
In such a setup, the Bell measurement can perform joint estimation with precision surpassing the bound established in Equation ~(\ref{rel1}) for separable measurements, as long as the amplitude of phase diffusion is less than $\Delta_0$, corresponding to $e^{-{\Delta_0}^2}=\sqrt{2/(1+\sqrt{5})}$. Indeed, for $\Delta=0$, a Bell measurement yields
\begin{equation}
\label{belltradeoff}
\frac{H_{1,1}^{-1}}{\text{Var}(\phi)}+\frac{H_{2,2}^{-1}}{\text{Var}(\Delta)}=\frac{3}{2}M,
\end{equation}
a value larger than the right side of the inequality given by Equation ~(\ref{rel1}), implying that greater precision can be obtained by investing in collective measurements.

For a larger value of phase diffusion, we perform a numerical search over all two qubit projective measurements that provides the achievable pairs of $\{\frac{H_{1,1}^{-1}}{M\,\text{Var}(\phi)},\frac{H_{2,2}^{-1}}{M\,\text{Var}(\Delta)}\}$, also optimizing for the smallest total of the entropy of entanglement for the corresponding projectors. Figure\ref{collectmeas} shows the results of this numerical search, revealing how a higher violation of the bound derived for separable measurements can be obtained with a more entangled measurement.

Our trade-off relations have been derived for those states whose evolution is effectively described in a 2D Hilbert space. In order to explore the form that this trade-off takes for probe state in a larger space, we present a numerical study of the performance of Holland-Burnett (HB) states \cite{HB}. An $\text{HB}(N)$ state results from the interference of two $N$-photon states on a beam splitter. $\text{HB}(N)$ states provide the same precision scaling as $\text{N}00\text{N}$ states, but are more resilient to losses than the latter.
\begin{figure}[H]
\centering
\includegraphics[width=0.475\textwidth]{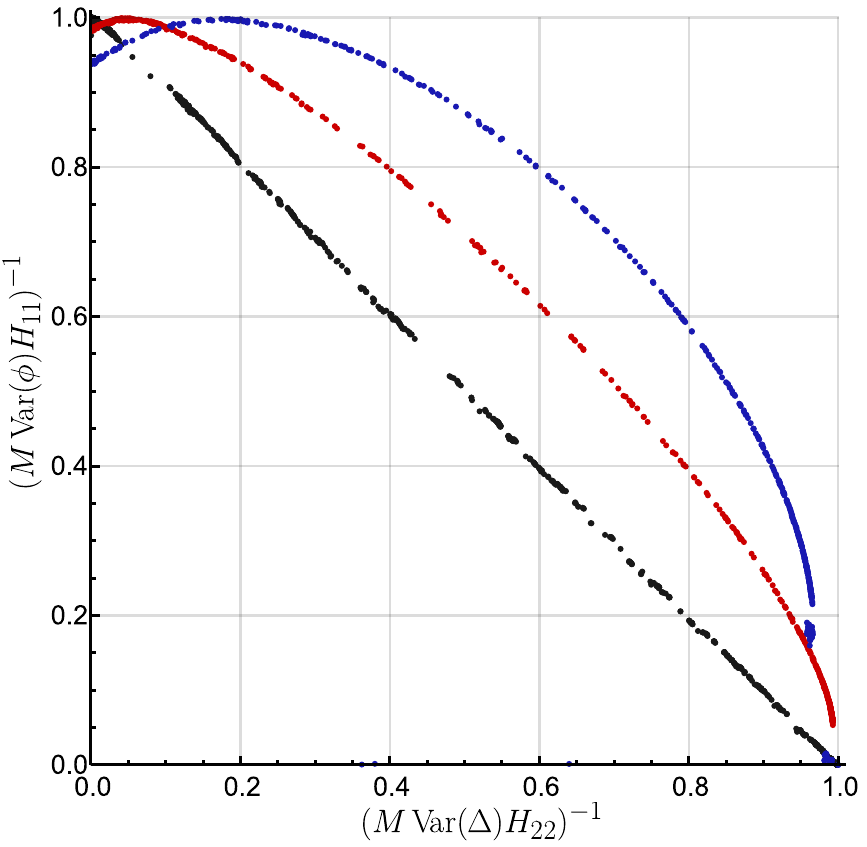}
\caption{\textbf{Joint estimation using HB states.}
Limits found by a simulated annealing search over all projectors acting on the space of the $\text{HB}(3)$ state. The black points correspond to $\Delta=0.01 \,\text{rad}$ ($\approx 0.6^\circ$), red points to $\Delta=0.05 \,\text{rad}$ and blue points to $\Delta=0.1 \,\text{rad}$. The details of the search are presented in the Methods section, in Supplementary Fig. 3 and in the Supplementary Methods.}
\label{HBstates}
\end{figure}
We performed a numerical search over all projective measurements on the 4D space corresponding to the $\text{HB}(3)$ state, optimizing the set of values $\{\frac{H_{1,1}^{-1}}{M\,\text{Var}(\phi)},\frac{H_{2,2}^{-1}}{M\,\text{Var}(\Delta)}\}$. The trade-off bounds observed in the results of the search depend on the amplitude of the phase diffusion. While for $\Delta= 0$, the linear trade-off expressed by Equation~(\ref{rel1}) is observed, for larger phase diffusion, we obtain limits higher than this (results are presented in Figure \ref{HBstates}). In Supplementary Fig. 4, we show how a photon number resolving measurement \cite{loss2} can beat the limit in Equation~(\ref{rel1}) when applied to HB states, however not reaching the bounds depicted in Figure \ref{HBstates}.

\begin{figure}[H]
\centering
\includegraphics[width=0.475\textwidth]{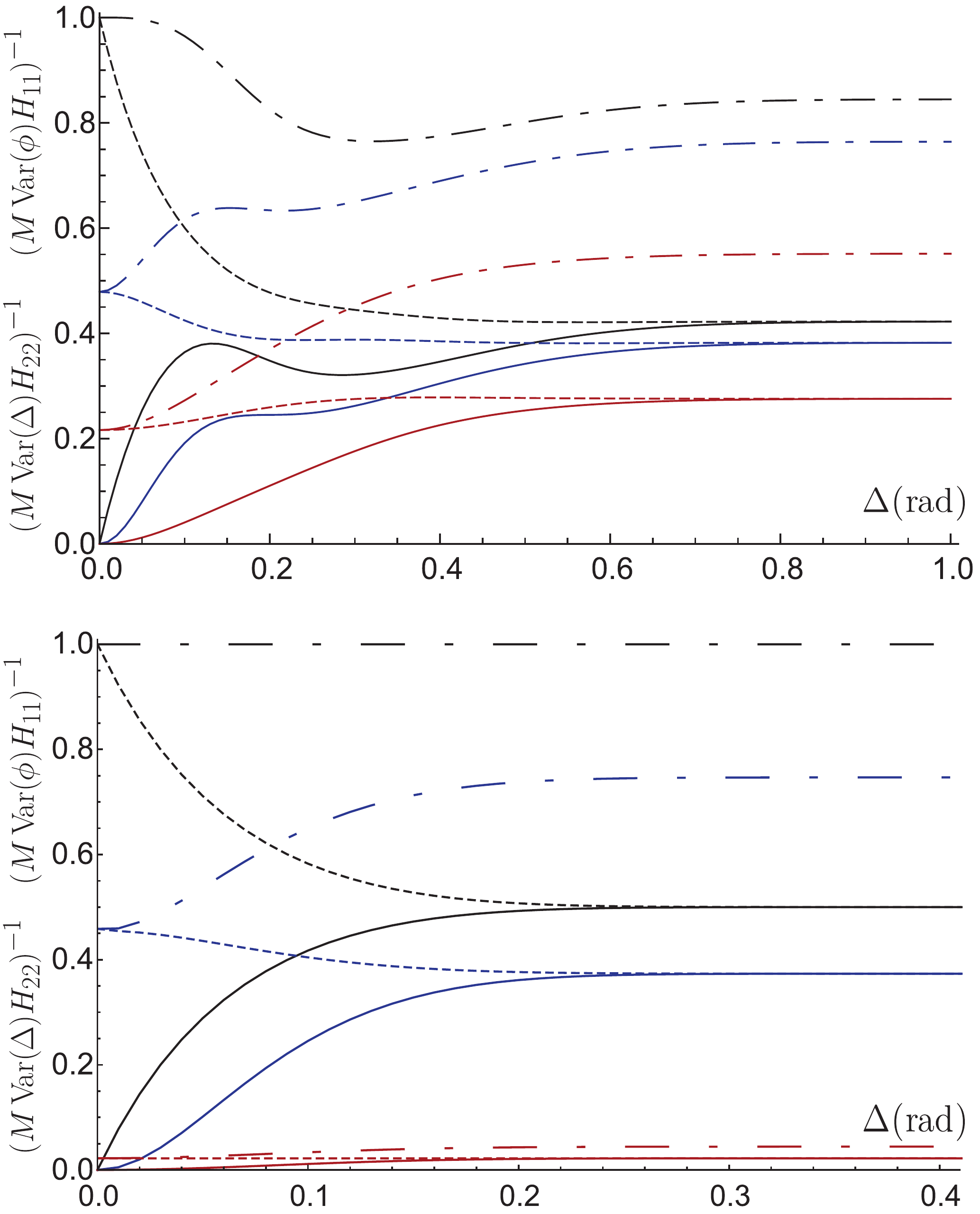}
\caption{\textbf{Practical setup with losses.}
The ratios between the optimum single parameter statistical variances and statistical variances that can be achieved with a double homodyne measurement; solid line for phase diffusion estimation, dashed line for phase estimation and dot-dashed line for the sum of the two; (a) for an $\text{HB}(3)$ probe state (with 6 photons) and (b) for a $\text{N}00\text{N}(6)$ probe state; with symmetric losses -- black for unit efficiency, blue for $0.95$ total efficiency and red for $0.5$ total efficiency. Note that the HB state is better suited for parameter estimation with loss than the $\text{N}00\text{N}$ state. 
}
\label{loss}
\end{figure}
It is recognized that the precision of any measurement making use of entangled states is affected by loss.  Here we illustrate a different effect of loss, i.e. how it affects the performance of simultaneous estimation. We focus on a practical scenario where the double homodyne measurement is used to analyze HB and $\text{N}00\text{N}$ states with 6 photons; the results are shown in Figure \ref{loss}. They illustrate the fact that HB states are more robust to loss than $\text{N}00\text{N}$ states not only in terms of QFI scaling, but also for attaining a satisfactory joint estimation precision.

We demonstrate in the Supplementary Methods that for all path-symmetric probe states (of which HB states are an example), with $\Delta=0$ and no loss, the double homodyne measurement estimates phase optimally \cite{parity1, parity2}. In our results (presented in Figure \ref{loss}), the decrease in sensitivity due to loss is partly contained in the decreasing value of the QFI. In addition, the classical Fisher information corresponding to double homodyne detection degrades with respect to the QFI due to the effect of the incoherent part of the loss-affected probe signal, which introduces noise in the measurement outcomes (this is detailed in Supplementary Fig. 5).

\section*{Discussion}



Figure \ref{exp} shows the variances that can be obtained in our experimental setup with a probe state that has a phase shift of 1$^\circ$ with respect to the optimal probe state. We highlight a somewhat overlooked aspect of parameter estimation: the sensitivity of the measurements to experimental imperfections in the alignment of the phase of the probe state. For a large extent of the settings of our tunable measurement, the precision of the estimates is robust to this small variation in phase. This is because the four-outcome POVM is capable of distinguishing between the rotation and the shrinking of the Bloch vector corresponding to the probe state. As we tune the measurement close to either of the extremal points, corresponding to $\{\sigma_x, \sigma_y\}$, this ability is compromised. While with a projective measurement ($\sigma_x$ or $\sigma_y$ for our setup), when there is no prior information on the amplitude of the phase diffusion, phase estimation is not possible, with a balanced setting of the weights given to pairs of projectors, our setup is tolerant to phase alignment (this is illustrated in Supplementary Fig. 6 and the Supplementary Discussion). Notably, the performance of the double homodyne setup described in this work is completely independent on the phase of the probe state.

We have applied our study to quantum correlated states that offer enhanced sensitivity for phase estimation. We have also shown that collective measurements can offer an advantage for joint estimation. However, entangled measurements in optics require either probabilistic schemes, which have limited applicability in metrology, or strong nonlinearities, which may be challenging, and at the edge of current technology.

We have found that states with correlations over multiple Fock layers, such as HB states, can perform better than $\text{N}00\text{N}$ states in terms of joint estimation.

\section*{Methods}

\subsection*{Experimental setup}

The source is a mode-locked Ti:sapphire laser, working in the pulsed regime, with central wavelength $830$nm, bandwidth of $32$nm and a repetition rate of $256\text{kHz}$. The preparation stage consists of a polarizing beam splitter (PBS1) which transmits only horizontally polarized light, followed by a half-wave plate (HWP1) and a quarter-waveplate (QWP1), used for the preparation of polarisation states for detector tomography. QWP2 is set at $45^{\circ}$, rotating $\ket{R}$ to $\ket{V}$ and $\ket{L}$ to $\ket{H}$. The displaced Sagnac interferometer consists of two slightly displaced counter-propagating modes of
equal length. The input state is split by PBS2 into its $\ket{H}$ and $\ket{V}$ components, corresponding to the two paths of the interferometer. After being acted upon by HWP2, the two paths recombine on PBS2. Depending on the orientation of HW2, the input beam is split and directed towards outputs $1$ and $2$. The polarisation state at output $1$ is approximately that of the input with a phase shift due to a path difference in the arms of the interferometer. QWP3 is a multiorder waveplate, with axis vertical, which is twisted in order to correct for this phase shift. The displaced Sagnac interferometer acts as a tunable non-polarizing beam splitter, with the added effect of switching $\ket{H}$ and $\ket{V}$ polarisations in output $2$. The detectors situated after HWP3 and PBS3 measure polarisations $\ket{D}$ and $\ket{A}$, respectively. The detectors situated after PBS4 measure $\{\ket{H}, \ket{V}\}$ in output $2$, effectively measuring part of the input polarisation state in the basis $\{\ket{R}, \ket{L}\}$. Single-mode fibers (SMF) are used to couple light into the detectors and alignment of the interferometer is performed by coupling the horizontal and vertical modes independently into SMF. The interferometer phase is set so that a minimum of interference is measured in output $1a$ when the input polarisation state is $\ket{D}$. The measured visibility of the interference was $\sim 97\%$.

We characterized the tunable measurement by performing detector tomography, with different settings of HWP3 and measuring intensities with a photodiode.

\subsection*{Estimation and errors}

The experimental errors affecting our setup fall into three categories: (1) statistical errors intrinsic to quantum measurement, which are the object of our study; (2) loss and distinguishability of photons, which are accounted for in the description of the setup and (3) technical (systematic) errors. The latter dominate statistical errors in our characterization of the setup. One of the two easy-to-identify error sources consists of intensity fluctuations on a time scale longer than the detection time, which can be dealt with by recording traces of the intensity readings and using the measured distributions when fitting data to the POVM model. The second consists of imperfections in the manufacturing and calibration of the waveplates used for preparation of the input polarisation state.

The POVMs are estimated by using a maximum likelihood algorithm comparing the collected data with the predictions from the reconstruction. We verify that the outcomes predicted by the reconstructed POVMs differ from those measured by values accountable for by observed fluctuations. The error bars for the estimated Fisher information are computed using a Monte Carlo simulation, starting with the variance of the measured values of light intensities. More detailed information on how Figure 3 was obtained is present in the Supplementary Methods.\\

\subsection*{Searches over projective measurements}

All elements of the set of projective measurements in a $d$-dimensional Hilbert space can be produced by acting on an orthonormal basis of this space with a unitary transformation. We perform simulated annealing \cite{optimization} over the set of projective measurements  by using random unitary transformations to perform a random walk. The algorithm decides whether a step is made in a randomly generated direction according to a tunable distribution that favours increasing values of $\frac{H_{1,1}^{-1}}{M\,\text{Var}(\phi)}$ and $\frac{H_{2,2}^{-1}}{M\,\text{Var}(\Delta)}$ and, for the search presented in Figure\ref{collectmeas}, decreasing values of the total entropy of entanglement of the projectors. We modify the step size, as well as the distribution controlling the random walk in order to reach the extreme values of the parameters that we are interested in, while ensuring that local minima are avoided. Details of this method, as well as arguments to restrict our search to projective measurements are presented in the Supplementary Methods.

\section*{Acknowledgments}
We thank Joshua Nunn, Tim Bartley, Mark Mitchison, David Jennings and Paolo Mataloni for discussion and comments. MV is supported by the EPSRC via the Controlled Quantum Dynamics CDT. XMJ is supported by an EU Marie-Curie Fellowship (PIIF-GA-2011-300820). WSK is supported by an EU Marie-Curie Fellowship (PIIF-GA-2011-331859). MGG acknowledges a fellowship support from
UK EPSRC (grant EP/I026436/1). MSK is supported by the Qatar National Research Fund (NPRP4-5554-1-084). This work was supported by the EPSRC (EP/H03031X/1, EP/K034480/1, EP/K026267/1), the European Commission project SIQS, the Air Force Office of Scientific Research (European Office of Aerospace Research and Development).

\section*{Author contributions}
M.D.V. developed the theory, designed and performed the experiment. M.G.G. contributed to the theory, G.D., X.-M. J. and W.S.K. contributed to the set-up and assisted with preliminary data collection, A.D. and M.B conceived and supervised the project and helped with designing the experimental setup, M.S.K. and I.A.W. supervised the project and contributed with discussion.

\bibliography{bibliography}
\end{document}